\documentclass{aa}
\usepackage{graphicx}
\begin{document}
   \title{Neutrinos from the pulsar wind nebulae}

   \author{W. Bednarek}

   \offprints{bednar@fizwe4.fic.uni.lodz.pl}

   \institute{Department of Experimental Physics, University of \L \'od\'z,
        ul. Pomorska 149/153, 90-236 \L \'od\'z, Poland
             }

   \date{Received ; accepted}

   \abstract{
In the recent paper we calculated the $\gamma$-ray spectra from pulsar wind nebulae
(PWNe), assuming that a significant amount of the pulsar rotational energy is converted 
into relativistic nuclei. These nuclei accelerate leptons which are responsible
for most of the observed electromagnetic emission from PWNe. Small part of nuclei 
also interact with the matter of the supernova producing $\gamma$-rays, which can also 
contribute to the observed spectra of young nebulae. Here  we 
calculate the spectra of neutrinos from the interaction of nuclei inside the nebula and 
the expected neutrino event rates 
in the 1 km$^2$ neutrino detector from: the Crab Nebula (PSR 0531+21), 
the Vela SNR (PSR 0833-45), G 343.1-2.3 (PSR 1706-44), 
MSH15-52 (PSR 1509-58), 3C58 (PSR J0205+6449), and CTB80 (PSR 1951+32).
It is shown that only the Crab Nebula can produce the neutrino event rate above the 
sensitivity limit of the 1 km$^2$ neutrino detector, provided that nuclei take most of 
the rotational energy lost by the pulsar. The neutrino event rate expected 
from the Vela SNR is comparable to that from the Crab Nebula but these neutrinos
are less energetic and emitted from a much larger region on the sky. Therefore it may be 
difficult to subtract the Vela SNR signal from the higher background of the atmospheric 
neutrinos. 

   }

   \maketitle
%

%
%
\section{Introduction}

The possibility that hadronic processes can contribute to the observed $\gamma$-ray 
emission from the PWNe has been considered in recent years by
e.g. Cheng et al. (1990) and Aharonian \& Atoyan~(1996). 
Following these works and earlier results (e.g. Berezinsky \& Prilutsky~1978),
we discussed a model for the production of radiation 
in hadronic processes with the application to the very young PWNe 
(Protheroe, Bednarek \& Luo~1998, Beall \& Bednarek~2002). In these papers,
the $\gamma$-ray and/or neutrino fluxes were estimated from the interaction of
nuclei, injected by the pulsar, with the matter and soft radiation of the supernova 
remnant during its early phase of development, i.e. within the  first few years.  
A general model like this has recently become very popular, finding its application in 
the estimation of neutrino fluxes from the $\gamma$-ray bursts in terms of 
the supranova model (e.g. Guetta \& Granot ~2002, Razzaque, Meszaros \& Waxman~2002, 
Dermer \& Atoyan~2003), which postulate the formation of a very fast pulsar during the
supernova explosion (Vietri \& Stella~1998). In another paper the contribution of nuclei 
from the pulsar to the $\gamma$-ray (and neutrino) spectrum was also calculated for 
the best known example of the older supernova remnant, i.e. the Crab Nebula 
(Bednarek \& Protheroe~1997). Recently, the neutrino flux from the Crab 
Nebula has been independently estimated by Amato, Guetta \& Blasi~(2003).
The authors predict the few to several neutrino events in a 1 km$^2$ detector per year 
from this source.

Recently we have considered a more general model for the hadronic and leptonic processes
inside the PWNe, which are based on the proposition that a significant part of the pulsar 
rotational energy is taken by relativistic nuclei (Arons and collaborators, e.g. 
Arons~1998). Our purpose was to calculate the $\gamma$-ray spectra
from the well-known PWNe. From the comparison of the calculations with the observations 
of these $\gamma$-ray nebulae, we derived some free parametrs of the considered model in 
order to predict the $\gamma$-ray and neutrino fluxes from other nebulae, which may become 
potential targets for the next generation 
of $\gamma$-ray and neutrino telescopes. The calculations of the $\gamma$-ray spectra 
from leptonic and hadronic processes
were reported in the accompanying paper by Bednarek \& Bartosik~(2003, BB03). The 
neutrino spectra are calculated in the present paper. 
Most of the earlier calculations of the $\gamma$-ray fluxes from the PWNe, which refer
mainly to the Crab Nebula (e.g. De Jager \& Harding~1992, Aharonian \& Atoyan~1995,
De Jager et al. 1996a, Hillas et al.~1998), were done based only on the leptonic 
origin for this emission. The calculations of the $\gamma$-ray fluxes from other nebulae 
are available only for a few objects, e.g. nebulae around PSR 1706-44 
(Aharonian, Atoyan \& Kifune~1997), PSR 1509-58 (Du Plessis et al.~1995), Vela pulsar
(De Jager et al.~1996b).

The neutrino fluxes from a few PWNe have recently been re-scaled from the observed 
TeV $\gamma$-ray fluxes by Guetta \& Amato~(2002), based on the simple assumption that 
in all these nebulae the $\gamma$-rays above 2 TeV are produced in a decay of $\pi^o$, 
which originate in 
hadronic processes. This assumption is in contradiction to the popular opinion on the 
relative importance of the leptonic and hadronic processes in PWNe and also with the 
results presented in this paper.

\section{The model for high energy processes in the pulsar wind nebula}

The details of the model for high energy processes in the PWNe are described in our first 
paper (BB03). For consistency we repeat here its main features.
The evolution of the supernova remnant, containing energetic pulsar, is described 
following the main points of the picture considered soon after the discovery of pulsars 
by  Ostriker \& Gunn~(1971) and Rees \& Gunn (1974). 
Let us assume that when the explosion occurs the expansion velocity of nebula 
at its inner radius is $V_{\rm 0,SN}$ and its initial mass is $M_{\rm 0,SN}$.
However this expension velocity can increase due to the additional supply 
of energy to the nebula by the pulsar and can also decrease due to the 
accumulation of the surrounding matter. We take these processes into account
when determining the radius of the nebula at the specific time, t, by using
the energy conservation,
\begin{eqnarray}
{{M_{\rm SN}(t)V_{\rm SN}^2(t)}\over{2}} = {{M_{\rm 0,SN}V_{\rm 0,SN}^2}\over{2}}
+ \int^t_0L_{\rm em}(t')dt',
\label{eq1}
\end{eqnarray}
\noindent
where 
\begin{eqnarray}
L_{\rm em}(t) = B_{\rm s}^2 R_{\rm s}^6 \Omega^4/6c^3\approx 
1.3\times 10^{45}B_{12}^2P_{\rm ms}^{-4}~~{\rm erg~s}^{-1}, 
\label{eq2}
\end{eqnarray}
\noindent
is the pulsar 
energy loss on emission of dipole electromagnetic radiation,
$\Omega = 2\pi/P$, and the period of the pulsar $P = 10^{-3}P_{\rm ms}$ s
changes with time according to 
\begin{eqnarray}
P^2_{\rm ms}(t) = P_{\rm 0,ms}^2 + 2\times 10^{-9}tB_{12}^2, 
\label{eq3}
\end{eqnarray}
\noindent
where $P_{\rm 0,ms}$ is the initial period of the pulsar and $B =10^{12}B_{12}$
G the strength of its surface magnetic field.
The expanding nebula increases the mass from the surrounding medium according to
\begin{eqnarray}
M_{\rm SN}(t) = M_{\rm 0,SN} + {{4}\over{3}}\pi \rho_{\rm sur} R_{\rm Neb}^3(t),
\label{eq4}
\end{eqnarray}
\noindent
where $\rho_{\rm sur}$ is the density of the surrounding medium and $R_{\rm Neb}$ 
is the outer radius of expanding envelope at the time, t, which depends on the 
expansion history of the nebula,
\begin{eqnarray}
R_{\rm Neb} = \int_0^t V_{\rm SN}(t')dt'.
\label{eq5}
\end{eqnarray}
The expansion velocity of the nebula, $V_{\rm SN}(t)$, and its
density of matter, $\rho_{\rm Neb} = 3M_{\rm SN}(t)/4\pi R_{\rm Neb}^3(t)$, at the time t
have been found by solving the above set of Eqs. (1-5) numerically. 

The pulsar loses energy in the form of relativistic wind which extends
up to the distance $R_{\rm sh}$. At this distance, the pressure of the wind is balanced 
by the pressure of the expanding nebula. We estimate the location of this shock 
at the time, t, by
comparing the wind energy flux, determined by $L_{\rm em}$ (Eq.~\ref{eq2}), with the 
pressure of the outer nebula, determined by the supply of energy to the 
nebula by the pulsar over the whole of its lifetime (Rees \& Gunn~1974),
\begin{eqnarray}
{{L_{\rm em}(t)}\over{4\pi R_{\rm sh}^2c}}\approx 
{{\int_0^tL_{\rm em}(t')dt'}\over{{{4}\over{3}}\pi R_{\rm Neb}^3}}.
\label{eq6}
\end{eqnarray}
\noindent
The assumption that the pulsar wind ram pressure is balanced by the total 
pressure inside the nebula is different from that made in the previous paper (BB03)
in which the wind pressure is balanced only by the magnetic field pressure. 
In fact this modification changes the location of the
pulsar wind shock for very young nebulae but do not have influence on the calculated 
neutrino fluxes from them since the escape conditions of heavy nuclei from the 
nebula do not change significantly.

Knowing how magnetic field depends on the distance from the pulsar in the pulsar
wind zone, we can estimate the strength of the magnetic field at the shock region
from 
\begin{eqnarray}
B_{\rm sh} = \sqrt{\sigma} B_{\rm pul}\left({{R_{\rm pul}}\over{R_{\rm lc}}}\right)^3
{{R_{\rm lc}}\over{R_{\rm sh}}},
\label{eq10}
\end{eqnarray}
\noindent
where $\sigma$ is the ratio of the magnetic energy flux to the particle energy 
flux from the pulsar at the location of the pulsar wind shock, 
$R_{\rm pul}$ and $B_{\rm pul}$ are the radius and the surface magnetic field 
of the pulsar, respectively. The evolution of $\sigma$ with the parameters of the pulsar 
is found 
by interpolating between the values estimated for the Crab pulsar, $\sim 0.003$, and
for the Vela pulsar, $\sim 1$ (see Eq. 16 and 
below in Bednarek \& Protheroe~2002 for details). 

It is likely that a significant part of the rotational energy of the pulsar, transfered
through the shock radius $R_{\rm sh}$ is in the form of relativistic heavy nuclei.
In fact, such an assumption can explain morphological features of the Crab Nebula and also
the appearance of extremely energetic leptons inside the nebula accelerated as a result 
of the rezonant scattering of positrons and electrons by heavy nuclei 
(Hoshino et al. 1992, Gallant \& Arons 1994). 
From the normalization to the observations of the Crab pulsar, Arons and collaborators
( e.g. see Arons~1998) postulate that the Lorenz factors of iron nuclei in the 
pulsar's wind should be
\begin{eqnarray}
\gamma_1\approx 0.3Ze\Phi_{\rm open}/m_{\rm i}c^2,
\label{eq20}
\end{eqnarray}
\noindent
where $m_{\rm i}$ and $Ze$ are the mass and charge of the iron nuclei, 
$c$ is the velocity of light, and $\Phi_{\rm open} = \sqrt{E_{\rm em}/c}$
is the total electric potential drop across the open magnetosphere.
Eq.~(8) postulates that the pulsar with a specific
period and a surface magnetic field accelerates nuclei monoenergetically.
As assumed in the Gallant \& Arons~(1994) model for the Crab Nebula, the nuclei 
take a significant part, $\chi$, of the total rotational energy lost by the pulsar.
The calculations of neutrino rates from the PWMe, presented below, are done for the
value of $\chi = 0.8$. In the next section we integrate the injection spectra of 
nuclei over the activity period of the pulsar in order to obtain the equilibrium 
spectrum of the nuclei inside the nebula at a specific time after pulsar formation.

\section{Production of neutrinos by nuclei}

\begin{figure}
  \vspace{5.5truecm}
\includegraphics{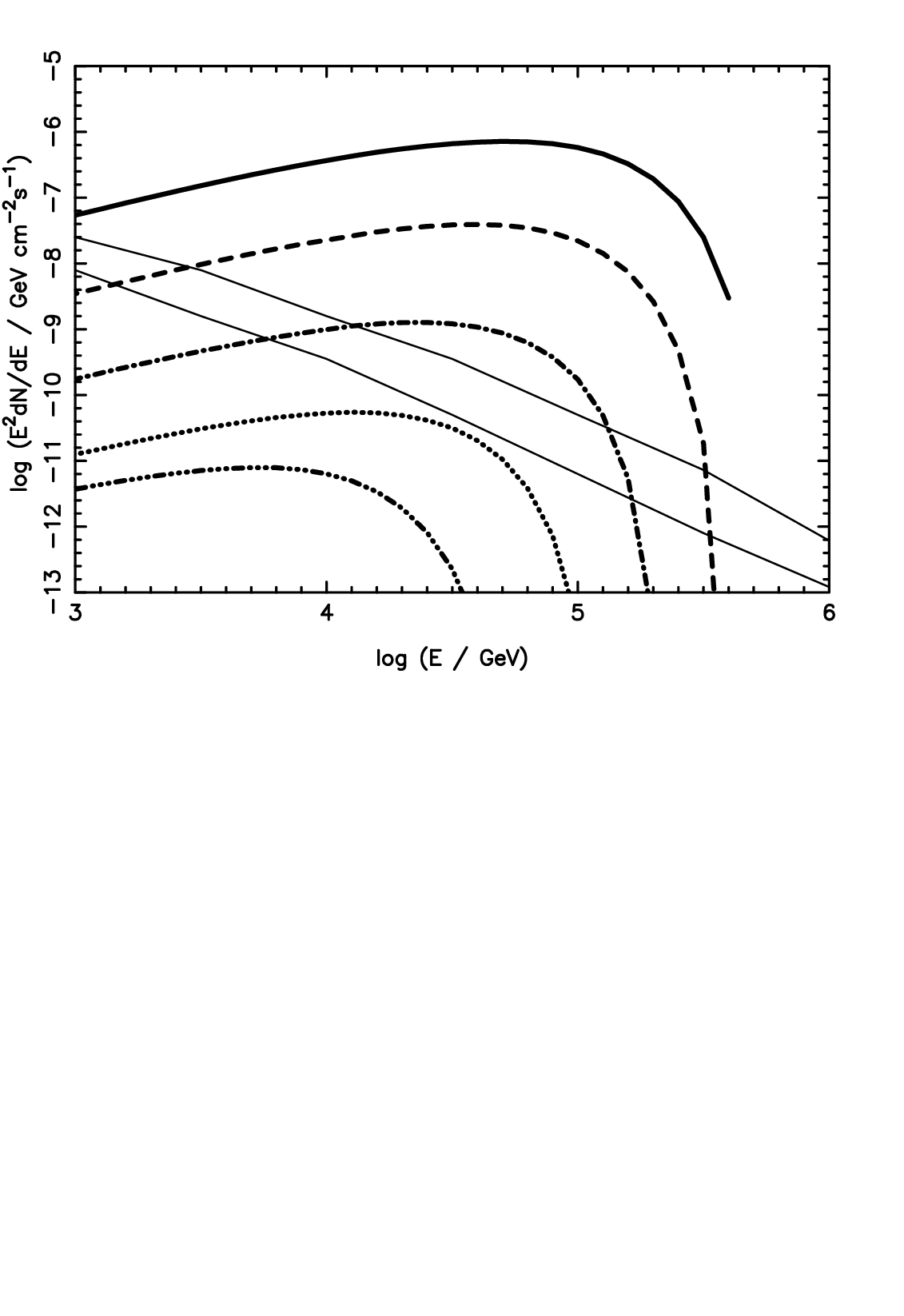}
  \caption{The muon neutrino and antyneutrino spectra from the pulsar wind nebula 
at the distance of 2 kpc at a different time after the pulsar formation:
$t = 10^2, 3\times 10^2, 10^3, 3\times 10^3$, and $10^4$ yrs. The initial period of the 
pulsar is 15 ms and the surface magnetic field $4\times 10^{12}$ G.
The supernova has the mass of 3 M$_\odot$ and the initial expansion velocity of
2000 km s$^{-1}$. The atmospheric neutrino background within $1^{\rm o}$ of the source, 
ANB, is marked by the thin full curves, horizontal (upper) and vertical (lower) 
(from Lipari 1993).}   
\label{fig1}
\end{figure}

The nuclei injected into the nebula suffer adiabatic energy losses, due to the 
expansion of the nebula, and rare collisions with the matter of the nebula. 
The most energetic nuclei can diffuse out of the nebula. 
We include all these energy loss processes for the nuclei (adiabatic losses, fragmentation, 
escape) following Bednarek \& Protheroe~(2002), and BB03, to
calculate the equilibrium spectra of different types of nuclei inside the nebula at 
the specific time after pulsar formation (see Sect. 4.1 and Fig. 2 in BB03
for example equilibrium spectra of nuclei inside the nebula).

The nuclei, with the equilibrium spectra, interact with the
matter of the supernova remnant and produce neutrinos via decay of pions. 
We estimate the neutrino spectra by simple re-scaling of the $\gamma$-ray spectra 
calculated for these hadrons, applying the scaling brake model proposed by Wdowczyk
\& Wolfendale (1987). The number of produced neutrinos is on average two times larger 
than the number of $\gamma$-rays (the multiplicity of charged pions in respect to the
neutral pions is larger by this factor) and the
average energies of neutrinos are a factor of two lower than the $\gamma$-rays.
As an example, we show in Fig. 1 the spectra of muon neutrinos (and antineutrinos) 
produced at a different times, $10^2$ yr, $3\times 10^2$ yr, 
$10^3$ yr, $3\times 10^3$ yr, and $10^4$ yr after the formation of the pulsar at the 
distance of 2 kpc. The initial 
parameters of the pulsar are: the period 15 ms and the surface magnetic field 
$4\times 10^{12}$ G, and the initial parameters of the nebula (supernova remnant) 
are: the mass $3 M_{\odot}$ and the velocity 2000 km s$^{-1}$. It is assumed that 
supernova exploded in the medium with a typical density of 0.3 cm$^{-3}$.
We also show the atmospheric neutrino background (ANB) expected within 1$^{\rm o}$ of
the source as calculated by Lipari (1993).
As expected, the intensities of neutrino spectra strongly depend on the age of the 
nebula due to the lower densities of matter inside the nebula.
The spectra also shift to lower energies due to the adiabatic energy losses of nuclei, 
more efficient escape of higher energy nuclei from the nebula, and lower energies of 
the freshly injected nuclei inside the nebula by older pulsars.
In Fig. 2 we show also the expected neutrino event rate in a 1 km$^2$ detector from pulsars
with different parameters. The predicted event rates are above 1 per year  if the 
age of the pulsar (and nebula) is shorter than $\sim 10^3 - 3\times 10^3$ yrs.

\begin{figure}
  \vspace{5.5truecm}
\includegraphics{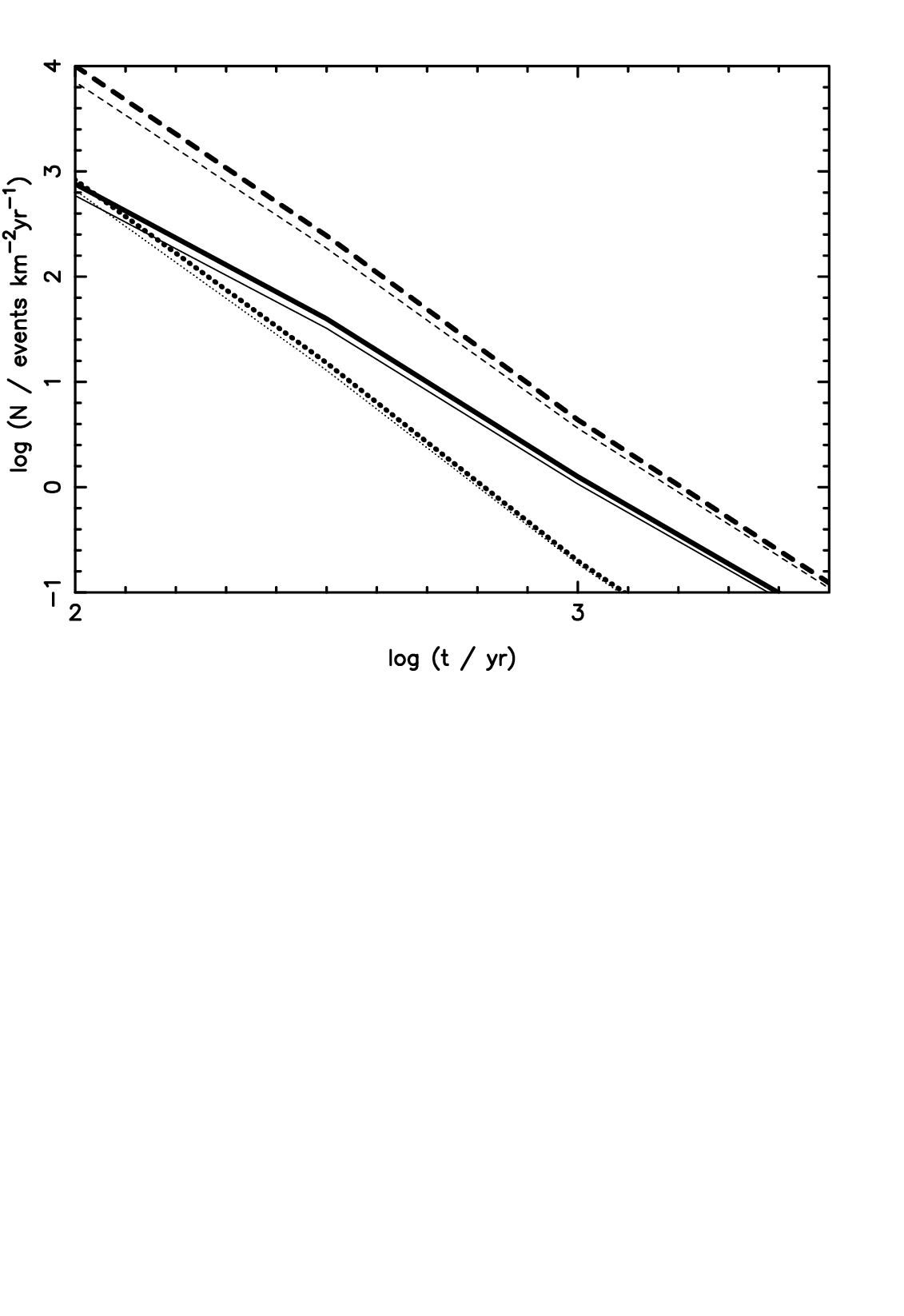}
  \caption{The muon neutrino event rate in 1 km$^2$ detector
from the pulsar wind nebula at the distance of 2 kpc as a function of the time after 
the pulsar formation for different initial parameters of the pulsar: initial period 15 ms 
and surface magnetic field $B = 4\times 10^{12}$ G (full curve),
5 ms and $B = 4\times 10^{12}$ G (dashed curve), and 15 ms and $B = 2\times 10^{13}$ G
(dotted curve). The rates for the neutrino events  from the nadir (absorption in the Earth) 
and the zenith directions are marked by the thin and thick curves, respectively.
}   
\label{fig2}
\end{figure}

\section{Expected neutrino rates from specific nebulae}

We calculate the expected neutrino spectra at the Earth and the neutrino event 
rates in a 1 km$^2$ 
detector from the PWNe for which the $\gamma$-ray fluxes have been calculated 
in our previous paper (BB03). The initial parameters of the nebulae and the pulsars for 
these sources have been taken as applied in calculations of $\gamma$-ray fluxes. 

To estimate the neutrino rates we applied the muon neutrino detection 
probabilities by the 1 km$^2$ detector calculated by Gaisser \& Grillo~(1987). 
The neutrinos arriving to the detector from the nadir direction can partially be absorbed 
by the Earth. This effect is significant for neutrinos with energies $> 100$ TeV. 
We take it into account by applying the neutrino absorption coefficients calculated by 
Gandhi~(2000). These absorption effects are not very important in the case of most of 
the considered objects.

\begin{figure}
  \vspace{5.5truecm}
\includegraphics{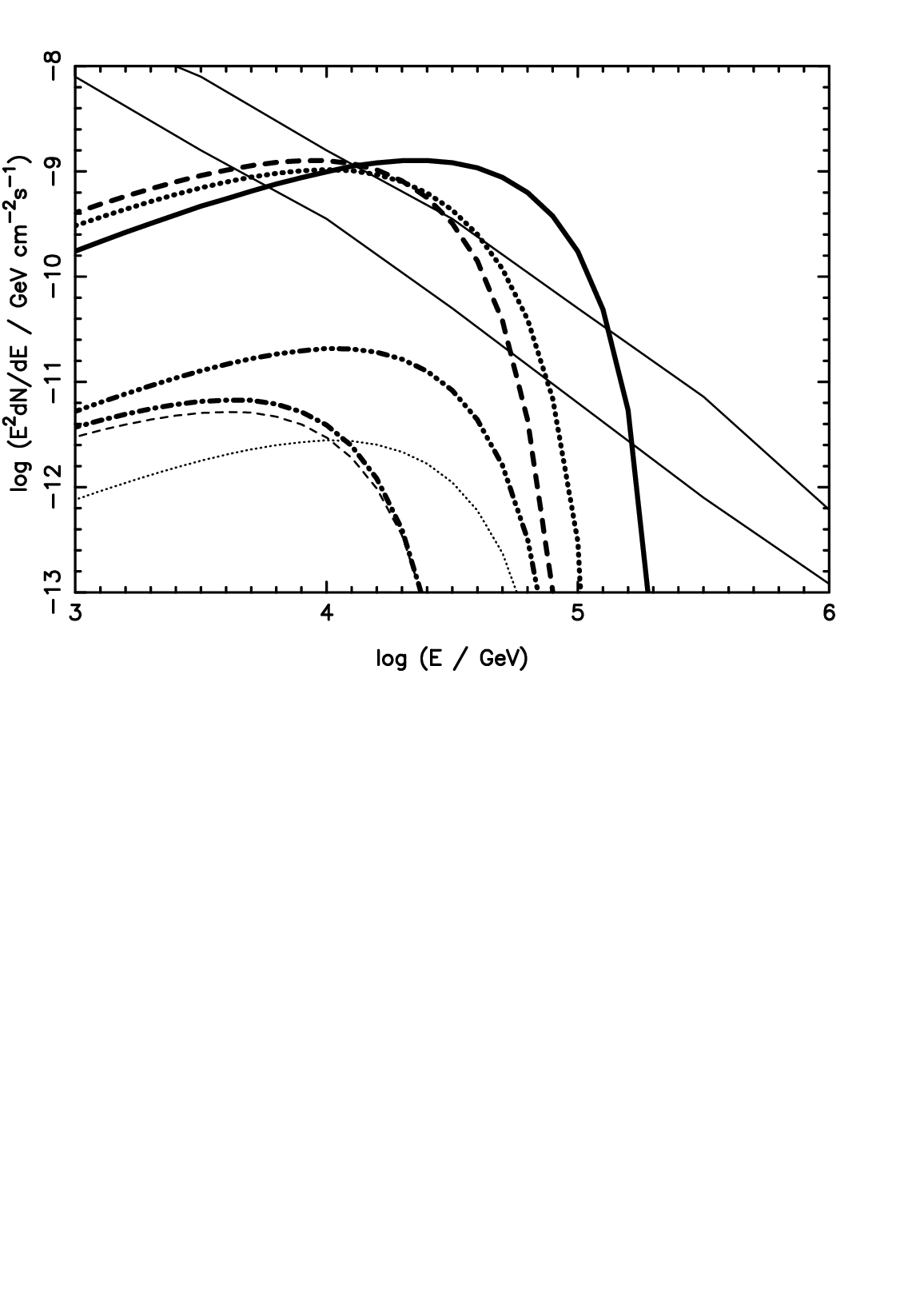}
  \caption{Neutrino spectra ($\nu_\mu$) expected from different nebulae: Crab Nebula - 
PSR 0531+21 (thick full curve),
Vela Nebula - PSR 0833-45 (thick dashed), G343.1-2.3 - PSR 1706-44 (thick dot-dashed), 
3C58 - PSR J0205+6449 (thick dot-dot-dot-dashed), MSH15-52 - PSR 1509-58 (thin dotted) 
and with close high density medium with 300 cm$^{-3}$ (thick dotted),
CTB80 - PSR 1951+32 (thin dashed). 
In all cases, except PSR 1509-58, the parent supernova explodes in 
the medium with the density of 0.3 cm$^{-3}$. The ANB is marked by the thin full curves.}   
\label{fig3}
\end{figure}
\subsection{The Crab Nebula}

The observed parameters of the Crab Nebula and the pulsar (the present
radius $\sim 2-3$ pc, the expansion velocity 2000 km s$^{-1}$, the period $\sim 33.4$ ms) 
can be fitted by assuming that:
(1) the pulsar was born with the initial period of 15 ms, the surface magnetic field
 $3.8\times 10^{12}$ G, as derived from the observed period and period derivative
with the assumption that the pulsar loses most of its energy on the dipole electromagnetic 
radiation; (2) the parent Crab Nebula supernova has
$3 M_{\odot}$ (consistent with the observed density of matter inside the nebula), and 
expends with the initial velocity of 2000 km s$^{-1}$ at the distance of 1830 pc from 
the Crab Nebula (Davidson \& Fesen~1985). For the applied parameters, we 
successfully described the observed $\gamma$-ray spectrum from the Crab Nebula (see BB03)
as a self-consistent composition of the $\gamma$-ray spectra produced by leptons 
($< 10$ TeV) and hadrons ($> 10$ TeV). 

The muon neutrino and antyneutrino spectrum from the Crab Nebula, calculated for the above 
parameters, is above the atmospheric neutrino background within $1^{\rm o}$  of the 
source 
(see Fig. 3). However the neutrino rate in the 1 km$^2$ detector, about one event per
year, is relatively low (see Table 1). This rate is similar to the neutrino rate
predicted in our earlier calculations (Bednarek \& Protheroe~1997), which are based on 
somewhat deferent assumptions. In that model
we assumed that neutrinos are produced in the interactions of nucleons with the matter 
of expanding supernova remnant. The nucleons were dissolved from the heavy nuclei during 
their propagation in the inner pulsar magnetosphere.

\subsection{The Vela Supernova Remnant}

The Vela pulsar is inside the compact nonthermal nebula with the radius of 
$\sim 7'$
and the extended Vela Supernova Remnant (Vela SNR) with the radius of $\sim 3.5^{\rm o}$. 
The observed parameters of the pulsar allow us to estimate its surface magnetic field, 
$4.5\times 10^{12}$ G, and the characteristic age of 11.300 yrs. We model the Vela SNR
assuming that the initial parameters of this object were similar to the Crab Nebula, 
i.e. the initial pulsar period of 15 ms, the mass of Vela SNR of 3 M$_\odot$ and the 
expansion velocity of 2000 km s$^{-1}$. The distance to the Vela pulsar 
is taken to be $\sim 300$ pc (Caraveo et al.~2001), although the older literature 
suggested the value of 500 pc (e.g. Cha, Sembach \& Danks~1999).
For these parameters of the pulsar and nebula, we estimate its
real age to be $\sim 6000$ yrs, consistent with the observed dimensions of the nebula. 
 
The muon neutrino spectrum, calculated from the Vela SNR, is just on the level of ANB 
within $1^{\rm o}$ around the source (Fig.~3). The expected neutrino event rate is similar 
to the Crab Nebula neutrino event rate (Table~1), but the neutrinos from the Vela SNR  
are less energetic. Moreover, due to the very large size of the Vela SNR on the sky, 
$\sim 3.5^{\rm o}$, the expected ANB for this source may be about an order of 
magnitude higher than marked in Fig.~3. Therefore the observation of a neutrino signal 
from the Vela SNR is unlikely.

\begin{table*} 
\caption{Expected number of $\nu_\mu$ neutrinos observed by the 1 km$^2$ 
detector during 1 year} 
\begin{tabular}{|c|c|c|c|c|c|c|}
\hline
Nebula & Crab Nebula & Vela nebula & G 343.1-2.3 & 3C58 & MSH15-52 & CTB80 \\
(pulsar) & (PSR0531+21) & (PSR0833-45) & (PSR1706-44) & (PSRJ0205+6449)  & 
(PSR1509-58) & (PSR1951+32) \\
\hline 
 & & & & & & \\
neutrino rate & 1.3 (1.1) & 1.15 (1.05) & $4\times 10^{-3}$  & 0.02 & $3\times 10^{-3}$
 & $5\times 10^{-3}$ \\   	    
($> 0.1$ TeV) & & & & & (1.0) & (0.03)\\
\hline 
\end{tabular} 
\label{tab1} 
\end{table*} 
\subsection{The Nebula around PSR 1706-44}

The pulsar PSR 1706-44, and its nebula (G 343.1-2.3), show close similarities to the 
Vela pulsar and SNR. It has also been reported at TeV $\gamma$-ray energies 
(Kifune et al.~1995, 
Chadwick et al.~1998). These pulsars have similar characteristic ages and present periods.   
However, due to the lower surface magnetic field of the PSR1706-44, equal to 
$3.1\times 10^{12}$ G, and a longer period, 102 ms, 
the real age of the pulsar has to be closer to its characteristic age 17.400 yrs.
As in the previous modelling we assume that the initial period of this pulsar is also 
15 ms and the expansion velocity of the nebula 2000 km s$^{-1}$.
For these parameters of the pulsar and nebula, the consistency with the observed parameters 
of this object is reached for its real age 16.000 yrs. The distance to PSR 1706-44 is 
taken 1.8 kpc (Taylor \& Cordes~1993).

The neutrino spectra calculated from PSR 1706-44 are about two orders of magnitude below 
the ANB (Fig.~3), and the predicted neutrino event rate in a 1 km$^2$ detector, 
$4\times 10^{-3}$ per year, do not allow their detection in a reasonable time.   
These small values are due to the significantly larger distance and higher real age of  
PSR 1706-44 in comparison to the Vela pulsar.

\subsection{The Nebula 3C58 around PSR J0205+6449}

The pulsar PSR J0205+6449 and its nebula 3C58 is sometimes identified  
with the historic supernova in 1181 yr (Thorsett et al.~1992). However, in such a case, 
using the required 
initial period of the pulsar of $\sim 60$ ms and the expension velocity of the nebula 
of $\sim 5000$ km s$^{-1}$ (to fit the observed dimensions of the nebula), we were not 
able to describe successfully 
the observed electromagnetic spectrum from this source in the broad energy range (BB03).

Therefore, in order to fit the observed electromagnetic spectrum, we followed another 
proposition for the parameters of this pulsar and nebula: the age of 5000 yrs 
(Bietenholz et al.~2001), which is consistent with the characteristic age 
(Murray et al.~2002), initial period of 15 ms, and expansion velocity of 1000 km s$^{-1}$ 
(Fesen 1983). These values allow to fit the 
present parameters of this system. The surface magnetic field of this pulsar is 
estimated to be $3.6\times 10^{12}$ G and its distance on 3.2 kpc (Roberts et al.~1993).
However for these parameters, the neutrino spectrum from PSR J0205+6449 is about two
orders of magnitude below the ANB (Fig. 3), and the predicted neutrino event rate 
(see Table 1) is not detectable.

\subsection{The Nebula MSH15-52 around PSR 1509-58}

The pulsar PSR 1509-58, with the characteristic age of $\sim 1700$ yr, and its complex 
nebula MSH15-52,
are sometimes identified with the historic supernova SN 185 (Thorsett 1992). The
distance to these objects is put in the range 4.2 kpc (kinematics of H I, Caswell et 
al. 1975) up to 
5.9 kpc (dispersion measure, Taylor \& Cordes 1993). We apply the value of 5.2 kpc.
The pulsar has the present period of $\sim 150$ ms, and the surface magnetic field 
of $1.5\times 10^{13}$ G. Its present parametrs can be explained by assuming that the 
initial pulsar period was 15 ms and its real age is close to its characteristic age. 
However, in order to fit the observed dimensions of the nebula, $5'\times 10'$, we have 
to assume that the expansion velocity of the bulk matter in this nebula is 5000 km 
s$^{-1}$, which is much higher than that applied in the modelling of other nebulae.

The neutrino spectrum calculated from this nebula is about two orders of magnitude below 
the ANB (Fig. ~3) and the expected neutrino event rate in 1 km$^2$ detector is very low
(Table~1). However this source has been marginally detected by the CANGAROO telescope
at TeV energies (Sako et al.~2000). We were not able to describe such a high level of the
$\gamma$-ray emission from MSH15-52 by only leptonic processes (BB03). However, the 
thermal, optical nebula RCW 89, containg $H_\alpha$ line-emitting filaments with the 
density of $\sim 5\times 10^3$ cm$^{-3}$, coincide with the NW component of 
the remnant MSH15-52 (Seward et al.~1983). Therefore, if the nuclei, injected by the 
pulsar are captured by these high density filaments, then
the reported level of $\gamma$-ray emission from this object can be explained
by the $\gamma$-rays produced in interactions of nuclei with the matter with average
density of $\sim 300$ cm$^{-3}$. Then, the accompanying neutrino emission is comparable to
the ANB (Fig. 3), and the expected neutrino event rate in a 1 km$^2$ detector is
close to 1 event per year.

\subsection{The Nebula CTB80 around PSR 1951+32}

The pulsar PSR 1951+32 has a short period, 39.5 ms, a large characteristic age of 
$1.1\times 10^5$ yr, and a relatively weak surface magnetic field of $4.9\times 10^{11}$ G.
It lies inside the supernova remnant CTB80, which consists of a $10'\times 6'$ compact 
nebula and shell-like extended component with a diameter of $30'$. The estimated dynamic 
age of the nebula $9.6\times 10^4$ yr (for the distance 2.5 kpc) matches the 
characteristic age of the pulsar (Koo et al.~1990). Therefore in our modelling, we assume
that these objects are at the distance of 2.5 kpc and have the real age close to the 
pulsar characteristic age. The observed parameters of the pulsar and nebula are consistent
with the initial period of the pulsar of 15 ms and the expansion velocity of the nebula
2000 km s$^{-1}$. The neutrino spectrum calculated for the above parameters is also 
orders of magnitudes below the ANB (Fig.~3), and the predicted neutrino event rate is 
below the sensitivity of a 1 km$^2$ neutrino detector (Table 1).

The nebula CTB80 is sometimes regarded as the remnant of the recent supernova in the year 
1408, found in Chinese records (Strom et al.~1980, Wang \& Seward~1984). If this is 
the case, then the initial period of the pulsar has to be very close to the one observed  
and the energy lost by the pulsar is very low. Assuming the 
expansion velocity of the nebula qeual to 1000 km s$^{-1}$, to fit the observed dimension
of the compact nebula, the expected neutrino event rate in this case is about an order of 
magnitude higher than estimated above (see the number in brackets in Table~1), but
still below the sensitivity of the future neutrino detectors.

\section{Conclusion}

According to the model discussed here for the high energy processes in PWNe, 
in which most of the observable $\gamma$-ray emission originates in leptonic processes 
and the accompanying hadronic processes contribute only to the higher energy part 
of the spectrum, only young PWNe (with the age 
$< 1-3\times 10^3$ yr and at the typical distance of 2 kpc) should be detected by a
1 km$^2$ neutrino detector with the rate above 1 event per yr, i.e. above the atmospheric
neutrino background within $1^{\rm o}$ of the source.
Between all considered PWNe, which are observed in the TeV $\gamma$-rays, contains a 
young 
$\gamma$-ray pulsar, or are considered as likely objects for TeV $\gamma$-ray detection,   
only neutrinos from the Crab Nebula should be observed by the 1 km$^2$ with the event 
rate of 1 per yr. The neutrino event rate from the Vela SNR is comparable to that of the 
Crab nebula, but due to significantly lower energies of neutrinos and a much larger 
solid angle of the source on the sky, the signal from this 
object may not be extracted from the atmospheric neutrino background. 
The neutrino spectra from other considered nebulae (pulsars):
G 343.1-2.3 (PSR 1706-44), MSH15-52 (PSR 1509-58), 3C58 (PSR J0205+6449), and 
CTB80 (PSR 1951+32), are significantly below the ANB, and the expected neutrino event
rate is far below the sensitivity of a 1 km$^2$ neutrino detector in a reasonable time
of observation. 

The neutrino rates, calculated in terms of the model for the high energy 
processes in PWNe considered here, are lower than those obtained 
by Guetta and Amato~(2002), who rescaled the observed $\gamma$-ray fluxes above 2 TeV
to the neutrino fluxes by simply assuming that all this $\gamma$-ray emission 
originates in hadronic processes. This assumption is in contradiction to the widely
accepted leptonic model for the origin of the bulk emission from the WPNe.
According to Guetta and Amato~(2002), Crab , Vela, PSR1706-44, and PSR1509-58, should be
easily detected by the 1 km$^2$ neutrino detector during one year.
We have calculated the ratio of the neutrino luminosity to the $\gamma$-ray 
luminosity above 1 TeV for the PWNe with reported TeV $\gamma$-ray emission. This ratio 
is equal to $L_\nu/L_\gamma\approx 58\%$ (the Crab Nebula), $22\%$ (the Vela Nebula), 
and $0.5\%$ 
(the nebula around PSR1707-44). Only in the case of the Crab Nebula the predicted neutrino 
luminosity should be close to the observed TeV $\gamma$-ray luminosity above 1 TeV. 
The neutrino fluxes can be comparable to the observed $\gamma$-ray fluxes even for 
relatively old nebulae if the high density 
medium is present close to the PWNa. As an example in Bednarek \& Bartosik~(2003), we 
consider the case of MSH15-52 (PSR 1509-58) which is close to the high density medium
(Seward et al.~1983). For this PWNa the ratio $L_\nu/L_\gamma$ is $\sim 77\%$.

In the case of all considered nebulae we assumed that the pulsars are born with the 
initial periods of 15 ms, as derived for the Crab pulsar. In fact, pulsars with shorter 
periods can also explain the main observed parameters of these nebulae. If the initial 
periods of these pulsars are significantly shorter then the expected neutrino event
rate increases. As an example, in Fig.~2 we consider the case of the pulsar with the 
period of 5 ms which produces the neutrino event rate by a factor of 2-3 higher than 
predicted for the 15 ms pulsar, for the nebulae older than $\sim 10^3$ yr. 
Therefore, significant reduction of the initial period of the pulsar does not 
influence the main conclusions of the paper concerning the detection of neutrinos
from specific nebulae by a 1 km$^2$ detector. 

We assumed that the density of a medium in which the supernova exploded is 
0.3 cm$^{-3}$. The pulsars surrounded by young supernova remnants, with similar 
parameters to the ones considered here, should produce much higher fluxes of neutrinos 
if they exploded in the high density medium.
For example, the presence of the cloud with the average density
of 300 particles cm$^{-3}$ close to MSH15-52 (PSR 1509-58) should allow detection of 
neutrinos and explain the marginal detection of TeV $\gamma$-rays from this object
(Sako et al.~2000), as a result of radiation produced during interactions of hadrons 
with the matter.

\begin{acknowledgements}
This work is supported by the Polish KBN grants No. 5P03D 025 21 
and PBZ-KBN-054/P03/2001. 
\end{acknowledgements}

\end{document}